\documentclass[prd,preprint,superscriptaddress,showpacs,nofootinbib,%
tightenlines
]{revtex4}
\usepackage{graphics}
\usepackage{epsfig}
\newcommand{\ben}{\begin{displaymath}}
\newcommand{\een}{\end{displaymath}}
\newcommand{\be}{\begin{equation}}
\newcommand{\ee}{\end{equation}}
\newcommand{\bea}{\begin{eqnarray}}
\newcommand{\eea}{\end{eqnarray}}

\begin{document}
\preprint{MKPH-T-03-11}
\title{Power counting in baryon chiral perturbation theory including vector
mesons}
\author{Thomas Fuchs}
\affiliation{Institut f\"ur Kernphysik, Johannes
Gutenberg-Universit\"at, D-55099 Mainz, Germany}
\author{Matthias R.~Schindler}
\affiliation{Institut f\"ur Kernphysik, Johannes
Gutenberg-Universit\"at, D-55099 Mainz, Germany}
\author{Jambul Gegelia}
\thanks{Alexander von Humboldt Research Fellow}
\affiliation{Institut f\"ur Kernphysik, Johannes
Gutenberg-Universit\"at, D-55099 Mainz, Germany}
\affiliation{High Energy Physics Institute,
Tbilisi State University,
University St.~9, 380086 Tbilisi, Georgia}
\author{Stefan Scherer}
\affiliation{Institut f\"ur Kernphysik, Johannes
Gutenberg-Universit\"at, D-55099 Mainz, Germany}
\begin{abstract}
   It is demonstrated that using a suitable renormalization condition one
obtains a consistent power counting in manifestly Lorentz-invariant baryon
chiral perturbation theory including vector mesons as explicit degrees of
freedom.
\end{abstract}
\pacs{
11.10.Gh,
12.39.Fe.
}
\date{August 1, 2003}
\maketitle

\section{\label{introduction} Introduction}
   Because of their phenomenological importance, vector mesons were
included in low-energy chiral Lagrangians already at an early stage
\cite{Schwinger:1967tc,Wess:1967jq,Weinberg:de}.
   Usually they were treated---within some approximation---as gauge bosons
of local chiral symmetry.
   Refs. \cite{Bando:1988br} and \cite{Meissner:1987ge} contain reviews of
these approaches including one of a ``hidden'' local chiral symmetry
(see also Ref.\ \cite{Harada:2003jx}).
   Details of the Lagrangian describing the interaction of vector
mesons with mesons and baryons in a chirally invariant way can be found
in Refs.\ \cite{Borasoy:1996ds,Ecker:yg,Ecker:1988te}.
  Different formulations of vector meson effective field theories were
shown to be equivalent in Ref.\ \cite{Birse:1996hd}.
   However, to the best of our knowledge the incorporation of
(axial) vector mesons into (baryon) chiral perturbation theory
\cite{Weinberg:1978kz,Gasser:1984yg,Gasser:1988rb}
remains an important open problem \cite{Meissner:2001fu} as long as a
systematic power counting is not available.
   In the present paper we show how the approach formulated recently in
Ref.\ \cite{Fuchs:2003qc} is capable to consistently include
(axial) vector mesons in the manifestly Lorentz-invariant formulation of
the effective field theory of the strong interactions.

   The basic idea of Ref.\ \cite{Fuchs:2003qc} (see also Refs.\
\cite{Gegelia:1999gf,Gegelia:1999qt}) can be summarized as follows.
   If one uses the modified minimal subtraction scheme
of (baryon) chiral perturbation theory ($\widetilde{\rm MS}$)
\cite{Gasser:1988rb}, then the diagrams
with an arbitrary number of loops contribute to lower-order calculations.
   As mentioned in Ref.\ \cite{Gasser:1988rb}, these contributions lead to a
renormalization of the low-energy constants, i.e., they can be absorbed
into a redefinition of these constants.
   The renormalized coupling constants of our extended on-mass-shell (EOMS)
scheme  correspond to a re-summation of the (infinite) series of
loop corrections.
   The coupling constants and fields of the $\widetilde{\rm MS}$ scheme
are expressed in terms of EOMS quantities.
   Expanding $\widetilde{\rm MS}$ quantities in terms of a power series
of EOMS-renormalized couplings generates counterterms
which precisely cancel the contributions of multi-loop diagrams to lower-order
calculations.
   The EOMS scheme thus leads to a consistent power counting in
baryon chiral perturbation theory.
   The inclusion of (axial) vector mesons into
this scheme does not introduce any new complications as long as
they appear only as internal lines in Feynman diagrams
involving soft external pions and nucleons with small three momenta.
   We will show this below by means of three select examples.

\section{\label{effective_lagrangian} Effective Lagrangian and Power
Counting}
   In our discussion we will make use of the effective
Lagrangian (including vector mesons) in the form given by Weinberg
\cite{Weinberg:de},
\begin{eqnarray}
{\cal L}&=&\frac{1}{2} \partial_\mu \pi^a\partial^\mu \pi^a
-\frac{M^{2}}{2}\pi^a \pi^a + \bar\Psi \left( i\gamma^\mu
\partial_\mu - m \right)\Psi -\frac {1}{4}F^a_{\mu\nu} F^{a
\mu\nu} +\frac {1}{2} M_\rho^2 A^a_\mu A^{a \mu}\nonumber\\
&&
+g \epsilon^{abc}\pi^a\partial_\mu\pi^bA^{c\mu}+g
\bar\Psi\gamma^\mu \frac{\tau^a}{2}\Psi A^a_\mu
- \frac{\stackrel{\circ}{g_A}}{2F} \bar\Psi\gamma^\mu \gamma_5 \tau^a \Psi
\partial_\mu \pi^a
+ {\cal L}_1,
\label{lagrangian}
\end{eqnarray}
where $\pi^a$ and $A^a_\mu$ are isotriplets of pion and $\rho$
meson fields with masses $M$ and $M_\rho$, respectively, and
$\Psi$ is an isodoublet of nucleon fields with mass $m$.
   The constants $F$ and $\stackrel{\circ}{g_A}$ denote the chiral limit of
the pion decay constant and the axial-vector coupling constant, respectively.
   Moreover, we use the universal $\rho$ coupling, i.e., $g=g_{\rho\pi\pi}=
g_{\rho NN}$.
   The field strengths are defined as
$F^a_{\mu\nu}=\partial_\mu A^a_\nu -\partial_\nu A^a_\mu
+g\epsilon^{abc} A^b_\mu A^c_\nu $ and ${\cal L}_1$ contains an infinite
number of terms.

   Our renormalization scheme is devised so that, after renormalization,
a given diagram has a so-called chiral power $D$ \cite{Weinberg:1991um}
which is determined by applying the following power counting rules.
   Let $Q$ collectively stand for the pion mass, a (small) external four
momentum of a pion or a (small) external three-momentum of a nucleon.
   The pion nucleon interaction counts as ${\cal O}(Q)$,
the $\rho$ nucleon interaction as ${\cal O}\left( Q^0\right)$,
the $\pi\pi\rho$ interaction as ${\cal O}\left( Q\right)$,
and the $\rho$ self-interaction terms as ${\cal O}\left( Q^0\right)$,
respectively.
   The integration of a loop counts as
${\cal O}\left( Q^n\right)$ (in $n$ dimensions).
   Finally, we count the $\rho$ meson propagator as
${\cal O}\left( Q^{0}\right)$, the nucleon propagator as
${\cal O}\left(Q^{-1}\right)$,\footnote{Fermion loops are integrated out, their
contributions being included in low-energy constants.} and the pion
propagator as ${\cal O}\left( Q^{-2}\right)$, respectively.

\section{\label{applications} Applications}
\begin{figure}
\epsfig{file=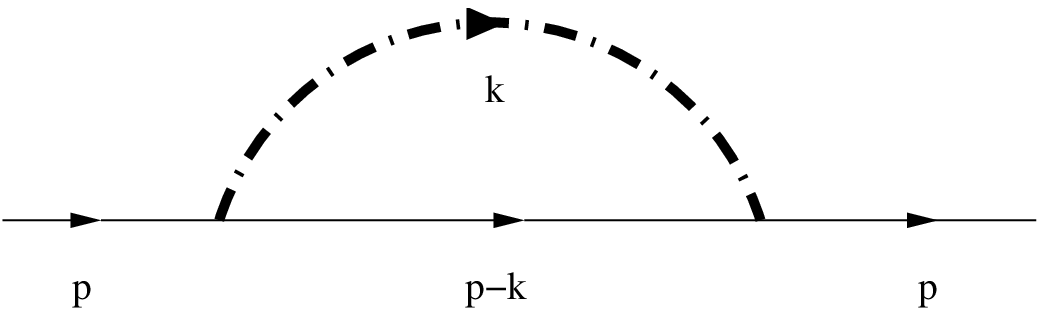,width=8cm}
\caption[]{\label{fse:fig}
One-loop contribution to the nucleon self energy due to
$\rho$ meson dressing.}
\end{figure}

   Let us now consider three characteristic diagrams illustrating our approach.
   We start with the vector-meson loop contribution to the nucleon
self energy (see Fig.\ \ref{fse:fig}):
\begin{equation}
\label{1lse}
-i \Sigma_1(p\hspace{-.45em}/)
= -\frac{g^2}{4}  \int \frac{d^nk}{(2\pi )^n}
\gamma_\nu\tau^a \frac{1}{p\hspace{-.45em}/
-k\hspace{-.45em}/ -m+i\epsilon}\gamma_\mu\tau^a
\frac{g^{\mu\nu}-\frac{k^\mu k^\nu}{M_\rho^2}}{k^2-M_\rho^2+i\epsilon}.
\end{equation}
   According to the above power counting $\Sigma_1$ is assigned the
chiral order $Q^{n-1}$.
   Calculating the expression of Eq.\ (\ref{1lse}) we obtain:
\begin{equation}
\label{1lsecalc}
\Sigma_1=-\frac{3 g^2}{8 M_\rho^2 p^2}
[AI_{\rho}(0)+B I_N(0)+C I_{\rho N}(0,-p)],
\end{equation}
where
\begin{eqnarray*}
A&=&p\hspace{-.45em}/\left[p^2-m^2+(2-n)M_\rho^2\right],\\
B&=&p\hspace{-.45em}/[p^2+m^2-(2-n)M_\rho^2]-2p^2 m,\\
C&=&p\hspace{-.45em}/[(3-n)M^2_\rho(p^2+m^2)-(2-n)M^4_\rho-(p^2-m^2)^2]
+2(n-1)M^2_\rho p^2 m,
\end{eqnarray*}
and
\begin{eqnarray}
I_\rho (0)&=&i \int \frac{d^nk}{(2\pi )^n}
\frac{1}{k^2-M_\rho^2+i\epsilon},\nonumber\\
I_N(0)&=&i \int \frac{d^nk}{(2\pi )^n}
\frac{1}{k^2-m^2+i\epsilon},\nonumber\\
I_{\rho N}(0,-p)&=&i \int \frac{d^nk}{(2\pi )^n}
\frac{1}{[k^2-M_\rho^2+i\epsilon] \left[
(k-p)^2-m^2+i\epsilon\right]}.
\label{1lintegrals}
\end{eqnarray}
   Before renormalization, the expression of Eq.\ (\ref{1lsecalc}) for
$\Sigma_1$ is of order $Q^0$, i.e., the unrenormalized diagram clearly
violates the power counting rules.

   To renormalize $\Sigma_1 $ we first apply the $\widetilde{\rm MS}$
scheme (modified minimal subtraction scheme of chiral perturbation
theory) \cite{Gasser:1984yg,Gasser:1988rb}.
   To obtain the final renormalized expression we perform additional finite
subtractions.
   For this purpose we expand the coefficients and integrals in
Eq.\ (\ref{1lsecalc}) in powers of $p^2-m^2$ and $p\hspace{-.45em}/-m$,
both counting as ${\cal O}(Q)$.
   The integrals $I_\rho(0)$ and $I_N(0)$ do not depend on $p^2$;
$I_{\rho N}(0,-p)$ is an analytic function of $p^2$ in the vicinity of
the point $p^2=m^2$.
   Hence the expansion of Eq.\ (\ref{1lsecalc}) in powers of $p^2-m^2$
contains only integer powers.
   The final renormalized expression for $\Sigma_1 $ is obtained by
subtracting from Eq.\ (\ref{1lsecalc}) the terms proportional to
$(p^2-m^2)^0$, $(p^2-m^2)$, $p\hspace{-.45em}/ -m$,
$(p^2-m^2)(p\hspace{-.45em}/ -m)$, and $(p^2-m^2)^2$.\footnote{We do not
quote the specific values of the subtraction constants, because they are
not relevant for our discussion.}
   As a result of this subtraction we obtain that the renormalized
self energy $\Sigma_1^R$ starts as $\left(p^2-m^2\right)^3$ in agreement with
power counting as $n\to 4$.
   All counterterms corresponding to the above subtractions are generated by
expanding bare quantities of the Lagrangian in terms of renormalized ones.

\begin{figure}
\epsfig{file=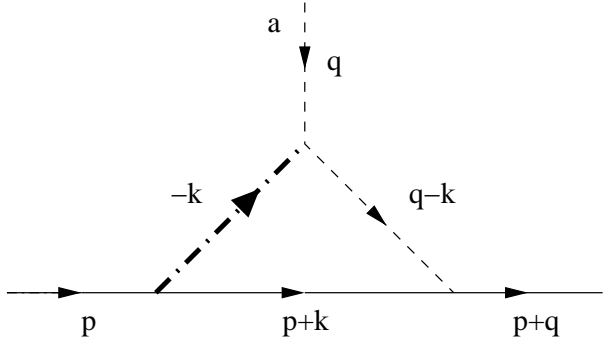,width=8cm}
\caption[]{\label{vertex:fig}
One-loop contribution to the $\pi NN$ vertex including an internal
$\rho$ meson.}
\end{figure}

   As the second example we consider the vertex diagram of Fig.\
\ref{vertex:fig}.
   Applying Feynman rules, the corresponding expression reads
\begin{equation}
V^a=\frac{i g^2 \stackrel{\circ}{g_A}}{2 F}\tau^a
\int \frac{d^nk}{(2\pi)^n}
(q\hspace{-.45em}/- k\hspace{-.45em}/)\gamma_5
\frac{1}{p\hspace{-.45em}/ +k \hspace{-.45em}/-m+i\epsilon}\gamma^\mu
\frac{g^{\mu\nu}-\frac{k^\mu k^\nu}{M_\rho^2}}{k^2-M_\rho^2+i\epsilon}
\,\frac{(2 q-k)^\nu }{(q-k)^2-M^2+i\epsilon}.
\label{vfd}
\end{equation}
   Power counting suggests that this diagram is of order $Q^3$.
   For the sake of simplicity, we only consider Eq.\ (\ref{vfd}) evaluated
between on-mass-shell nucleons [$p^2=m^2=(p+q)^2$],\footnote{
Recall that $\bar{u}(p+q)\gamma_5 u(p)$ counts as ${\cal O}(Q)$
\cite{Gasser:1988rb}.}
\begin{eqnarray}
\label{vfdc}
\bar{u}(p+q)V^a u(p)&=&g^2 \stackrel{\circ}{g_A}\frac{m}{F}
\{A\,I_{\rho\pi}(0,-q)+B\,[I_{\rho\pi}(0,-q)-I_{\rho\pi}(0,0)]
+C\, J(112|n+2)\nonumber\\
&&+D\, J(121|n+2)
+E\, I_{\rho\pi N}(0,-q,p)
+F\, I_\rho(0)\}\bar{u}(p+q)\gamma_5\tau^a u(p),
\end{eqnarray}
where the coefficients $A,\cdots, F$ are given by
\begin{eqnarray*}
A&=&-\frac{1}{2}\left( 5+\frac{q^2-M^2}{M_\rho^2}\right),\\
B&=&\frac{M_\rho^2-M^2}{2 q^2}\left(1+\frac{q^2-M^2}{M_\rho^2}\right),\\
C&=&8\pi(2m^2-q^2),\\
D&=&-8\pi q^2,\\
E&=&-2q^2,\\
F&=&\frac{1}{M_\rho^2},
\end{eqnarray*}
   and the integrals read
\begin{eqnarray}
\label{rpndef}
I_{\rho \pi}(0,p)&=&i \int \frac{d^nk}{(2\pi )^n}
\frac{1}{(k^2-M_\rho^2+i\epsilon)[(k+p)^2-M^2+i\epsilon]},\nonumber\\
I_{\rho \pi N}(0,-q,p)&=&i \int \frac{d^nk}{(2\pi )^n}
\frac{1}{(k^2-M_\rho^2+i\epsilon)[(k-q)^2-M^2+i\epsilon]
[(k+p)^2-m^2+i\epsilon]}.
\end{eqnarray}
   Moreover, we have introduced the auxiliary integral
\begin{equation}
\label{vfdint}
J(abc|d)=i\int \frac{d^dk}{(2\pi)^d}\frac{1}{
(k^2-M_\rho^2+i\epsilon)^a [ (k-q)^2-M^2+i\epsilon]^b
(k^2+2 p\cdot k+i\epsilon)^c},
\end{equation}
which, with shifted space-time dimension $n+2$, contributes to
Eq.~(\ref{vfdc}) as a result of reducing tensor integrals to scalar
ones.
   An explicit calculation yields
\begin{eqnarray}
I_{\rho\pi}(0,-q)&=&\frac{-1}{ (4 \pi )^{\frac{n}{2}}}\left(
M_\rho^2\right)^{\frac{n}{2}-2}\sum_{l=0}^\infty \frac{1}{l!}
\left( \frac{q^2}{M_\rho^2}\right)^l\frac{\Gamma \left(
2-n/2+l\right)\Gamma \left( 1+l\right)\Gamma \left(
1+l\right)}{\Gamma \left( 2+2 l\right)}\nonumber\\ &&\times
F\left( 1+l,2-\frac{n}{2}+l; 2+2 l;
1-\frac{M^2}{M_\rho^2}\right)\nonumber\\ &=&-\frac{1}{ (4
\pi)^{\frac{n}{2}}} \left( M^2_\rho\right)^{\frac{n}{2}-2}
\Gamma\left(2-n/2\right) -\frac{1}{16 \pi^2}
-\frac{1}{8\pi^2}\frac{M^2}{M_\rho^2}\ln\left(\frac{M}{M_\rho}\right)
-\frac{q^2}{32\pi^2 M_\rho^2}\nonumber\\ &&+{\cal
O}\left(Q^4\right), \label{vfdintexpl}
\end{eqnarray}
where $F(a,b;c;z)$ is the hypergeometric function \cite{Abramowitz} and
$Q$ stands for either $q$ or $M$.

   Again, we find that the unrenormalized diagram violates the
power counting.
   We renormalize Eq.~(\ref{vfdc}) by first subtracting
all ultraviolet divergences using the $\widetilde{\rm MS}$ scheme
which amounts to dropping all terms proportional to $\bar{\lambda}$
[see Eqs.\ (\ref{vfdcsub}) and (\ref{lambdabarapp}) below].
   To determine the additional finite subtractions, we expand all
coefficients and integrals in powers of $M^2$ and $q^2$.
   The integrals contain non-analytic parts which are proportional to
non-integer powers of $M^2$ and/or $q^2$ for non-integer $n$.
   These non-analytic contributions separately satisfy the power counting.
   We only expand the analytic parts and obtain the final
renormalized expression of the diagram by subtracting all terms of
the above expansion which are of order $Q^2$ or less.
   The terms which need to be subtracted read
\begin{eqnarray}
\bar{u}(p+q)V_{\rm subtr}^a u(p)&=&g^2
\stackrel{\circ}{g_A}\frac{m}{2F} \left\{\frac{5
(M^2_\rho)^{\frac{n}{2}-2}}{(4\pi)^\frac{n}{2}} \
\Gamma\left(2-n/2\right)+\frac{5}{16\pi^2}-\frac{1}{32\pi^2}
\right.\nonumber\\ &&\left.+32\pi m^2 J_0(112|n+2)
+\frac{2}{M_\rho^2}I_\rho(0)
-2\,\frac{q^2-M^2}{M_\rho^2}\bar{\lambda}
\right\}\bar{u}(p+q)\gamma_5\tau^a u(p),\nonumber\\
\label{vfdcsub}
\end{eqnarray}
where
\begin{equation}
J_0(abc|d)=i \int \frac{d^dk}{(2\pi)^d}
\frac{1}{(k^2-M_\rho^2+i\epsilon)^a
(k^2+i\epsilon)^b
(k^2+2 p\cdot k+i\epsilon)^c},
\label{vfdint0}
\end{equation}
and
\begin{equation}
\bar\lambda ={m^{n-4}\over 16\pi^2}\left\{ {1\over n-4}-{1\over 2}
\left[ \ln (4\pi) +\Gamma '(1)+1\right]\right\}.
\label{lambdabarapp}
\end{equation}
   Taking into account that
\begin{displaymath}
J(112|n+2)-J_0(112|n+2)\sim Q^2,
\end{displaymath}
it is now straightforward to check that the difference of Eq.\ (\ref{vfdc})
and Eq.\ (\ref{vfdcsub}) satisfies the power counting, i.e., is of
order $Q^3$ as $n\to 4$.

\begin{figure}
\epsfig{file=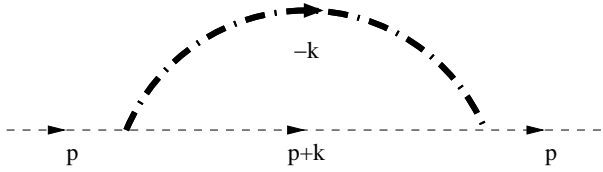,width=8cm}
\caption[]{\label{pise:fig} Pion self energy diagram with $\rho$ meson
dressing.}
\end{figure}

   As a final example, we discuss the one-loop diagram of the
pion self energy given in Fig.~\ref{pise:fig}.
   The corresponding expression reads
\begin{equation}
-i \Sigma^{ab}_\pi(p) =
-i \Sigma_\pi(p)\delta^{ab}=- g^2 \epsilon^{acd}\epsilon^{bcd} \int
\frac{d^nk}{(2 \pi )^n}  \frac{(2p+k)^\mu (2p+k)^\nu \left(
g_{\mu\nu}-\frac{k_{\mu}k_{\nu}}{M_{\rho}^2}\right)}{[
(k+p)^2-M^2+i\epsilon](k^2-M_{\rho}^2+i\epsilon)},
\label{pionse}
\end{equation}
from which we obtain
\begin{equation}
\Sigma_\pi(p) = -2 g^2  \left\{
\left(1+\frac{p^2-M^2}{M_{\rho}^2}\right)
I_{\rho}(0)+\left[ 2 M^2+2
p^2-M_\rho^2-\frac{\left( p^2-M^2
\right)^2}{M_\rho^2}\right]I_{\rho\pi}(0,p)-I_\pi(0) \right\},
\label{pionsecalc}
\end{equation}
where
\begin{equation}
I_{\pi}(0)=-\frac{ M^2\left( M_{\rho}^2\right)^{\frac{n}{2}-2}\Gamma
(2-n/2)}{(4 \pi)^{\frac{n}{2}}}-\frac{M^2}{16 \pi^2}+\frac{M^2}{8
\pi^2}\ln\left(\frac{M}{M_{\rho}}\right),
\label{deltapi}
\end{equation}
and $I_{\rho\pi}(0,p)$ is given in Eq.\ (\ref{vfdintexpl}).

   To renormalize the pion self energy we first apply the
$\widetilde{\rm MS}$ scheme.
   The additional finite subtraction counterterms are obtained by
expanding the coefficients and the analytic parts of the integrals
in Eq.~(\ref{pionsecalc}) and identifying those terms
which are of lower order than suggested by power counting, i.e.,
order $Q^4$.
   Note that the last term in Eq.~(\ref{deltapi}),
which is non-analytic in $M$,
will give a contribution to Eq.~(\ref{pionsecalc}) which, if taken
separately, violates the power counting.
   It cannot be removed by counterterms in the Lagrangian, but exactly
cancels with an analogous contribution coming from the $I_{\rho\pi}(0,p)$
term in Eq.\ (\ref{pionsecalc}).
   We arrive at the following renormalized expression:
\begin{eqnarray}
\Sigma^R_\pi(p)&=&- 2 g^2 \left\{ (2 M^2+2 p^2-M_\rho^2)
\left[ I_{\rho\pi}(0,p)+\frac{1}{16
\pi^2}+\frac{ \left( M_\rho^2\right)^{\frac{n}{2}-2}\Gamma \left(
2-\frac{n}{2}\right)}{(4 \pi )^{n/2}} \right]\right.\nonumber\\
&&\left.
-\frac{\left( p^2-M^2 \right)^2}{M_\rho^2} \left[
I_{\rho\pi}(0,p)-2 \bar\lambda \right] -\frac{ p^2}{32
\pi^2}-\frac{M^2}{8 \pi^2}\ln\left(\frac{M}{M_\rho}\right)\right\}.
\label{pionseren}
\end{eqnarray}
   Using Eq.\ (\ref{vfdintexpl}) we see that
Eq.\ (\ref{pionseren}) satisfies the power counting, i.e., is of
order $Q^4$.

   Two- (and multi-) loop diagrams have a more complicated structure,
but the outcome remains the same.
   Those terms which are non-analytic in small expansion parameters
satisfy the systematic power counting after subtracting the one-loop-order
sub-diagrams.
   The contributions which violate the power counting are analytic in
small expansion parameters and are subtracted by a finite number of
local counterterms in the Lagrangian.

\section{\label{summary} Summary and Conclusions}
   We have demonstrated that the inclusion of explicit
degrees of freedom corresponding to (axial) vector particles in
manifestly Lorentz-invariant baryon chiral perturbation theory does
not violate the power counting if a suitable renormalization
condition is used.
   As an important test of our method it is now necessary to apply a
full calculation to physical processes such as, e.g., the determination
of the nucleon electromagnetic form factors \cite{Schindler}, where a
one-loop calculation in ordinary baryon chiral perturbation theory does
not show a satisfactory agreement with data beyond very small values of
$Q^2\approx 0.1\,\mbox{GeV}^2$ \cite{Fuchs:2003ir,Kubis:2001zd}.

\acknowledgments
   The work of T.F.~and S.S.~was supported by the Deutsche
Forschungsgemeinschaft (SFB 443).
   J.G.~acknowledges the support of the Alexander von Humboldt Foundation.

\newpage

\end{document}